\documentclass[slac_one]{revtex4}
\usepackage{graphicx}
\usepackage{fancyhdr}
\begin{document}

\title{Early Top Physics at CMS Experiment} 

%

\author{J. M. Vizan}
\affiliation{Universidad de Oviedo, Spain}

\begin{abstract}
The top quark was discovered at the Tevatron in 1995. For the last
decade the study of its properties has been a major theme in the
worldwide experimental high energy physics program. The advent of
the LHC opens up a new era in top quark physics; because of the
large $t\bar{t}$ cross-section and the high luminosity, the LHC can
be thought of as a top factory. Here we present the prospects and
plans for ttbar physics at CMS at an early stage of the experiment,
covering from the initial establishment of the top signal, to the
first measurements that become possible for an integrated luminosity
of 100 $pb^{-1}$, considering a realistic detector performance.
\end{abstract}

\maketitle

\thispagestyle{fancy}


\section{INTRODUCTION} 
At the Large Hadron Collider (LHC) top quark pairs ($t\bar{t}$) will
be produced copiously. The next-to-leading order cross section at
LHC energies has been estimated to be 833 pb~\cite{refttcross}. Due
to this large cross section, at fairly low integrated luminosities
ranging from approximately 10 $pb^{-1}$ to 100 $pb^{-1}$ of LHC
data, a top quark signal could be established and the first top
quark pair production cross section measurements could be performed.
More details are provided in section 2. In addition, top quark
events can be used for the calibration of residual jet energy scale
corrections, as it is presented in section 3, or b-jet
identification efficiencies~\cite{refbtag}.

For all samples used in the different studies presented here $pp$
collisions were generated at $\sqrt{s} = 14$ TeV. In order to mimic
the realistic conditions during the early data taking, effects due
to tracker misalignment and calorimeter miscalibration were taken
into account in the reconstruction. The effect of multiple
proton-proton interactions was not included.

\section{TOP QUARK OBSERVATION AND CROSS SECTION MEASUREMENTS}

\subsection{Introduction}
For the different analysis reviewed in this section the event
selection strategy is based on a sequential cut procedure based on
robust variables. Among them, lepton and jet transverse momentum or
energy, lepton isolation variables, electron identification related
variables, or missing transverse energy. All relevant backgrounds,
except the single top production have been taken into account.
Different sources of systematic errors have been considered,
including those associated to lepton energy scale, trigger
efficiencies, jet reconstruction and energy scale, missing
transverse energy, parton distribution functions and background
cross sections.

The observability and cross section measurements of the $t\bar{t}$
production are reviewed in the following subsections for different
possible final states of the $t\bar{t}$ decay. The semileptonic muon
channel, the dilepton channel and the $e\tau$ and $\mu\tau$ channels
are considered in sections 2.2, 2.3 and 2.4 respectively. More
details are provided
in~\cite{refsemi},~\cite{refdilexp},~\cite{refdilmea}
and~\cite{reftau}.

\subsection{Observability of top quark pair production in the semileptonic muon channel with the first 10 $pb^{-1}$ of CMS data}
\label{secsemi} 
\begin{figure*}[t]
\centering
\includegraphics[width=60mm]{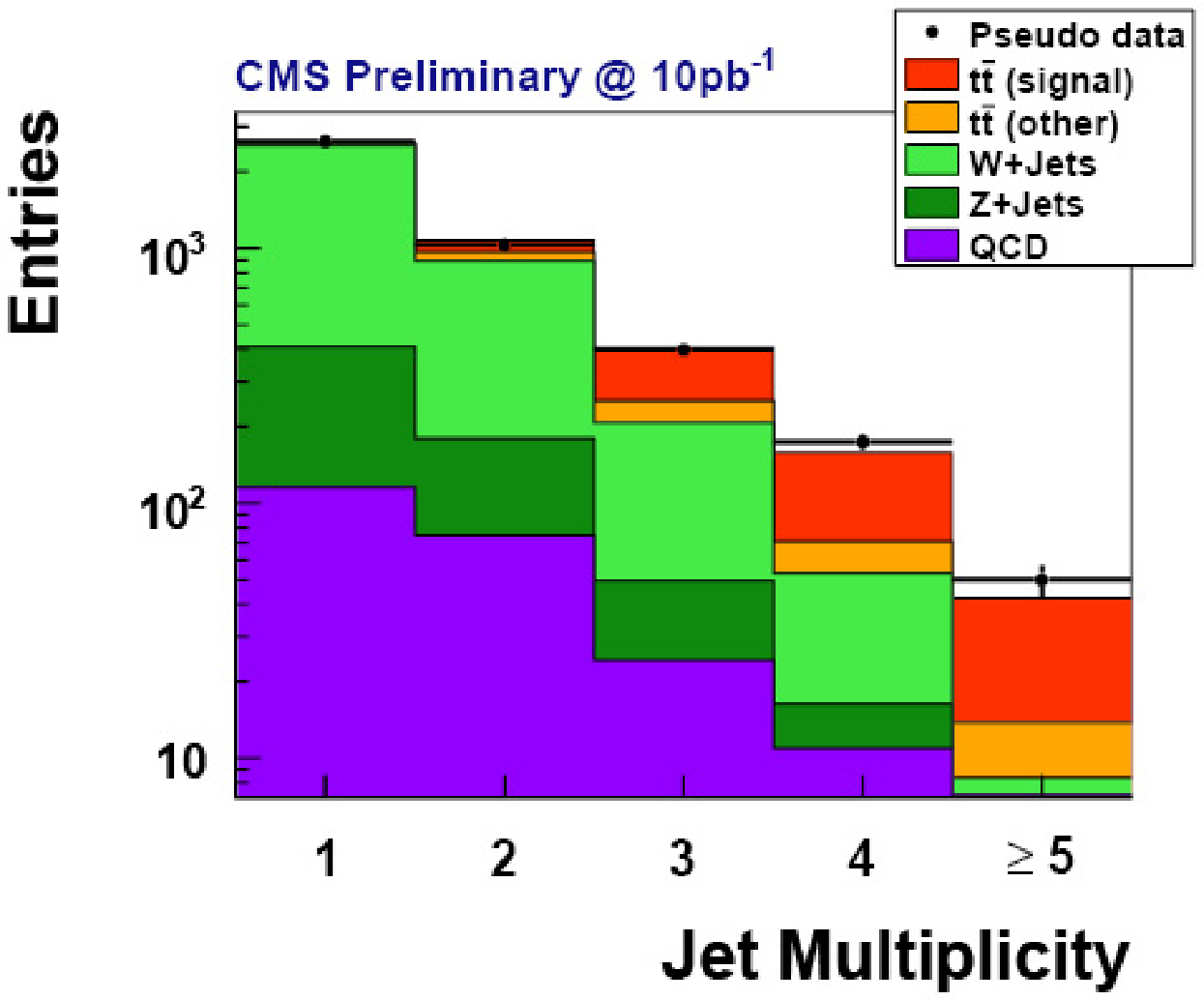}
\includegraphics[width=60mm]{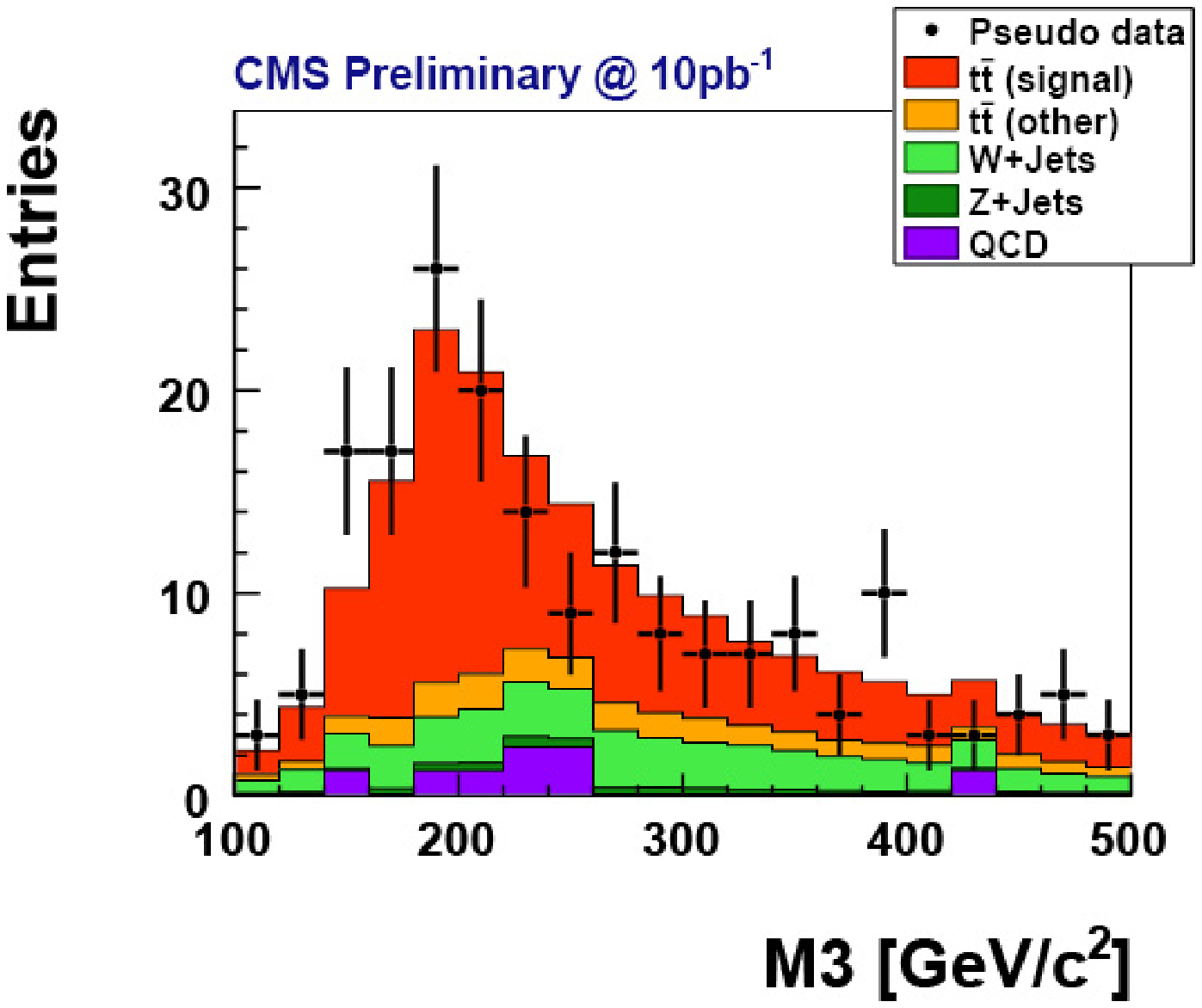}
\caption{Jet multiplicity distribution for events passing the final
selection except the requirement of $N_{jets} \geq{4}$ (left);
Invariant mass of the three jets with the highest vectorially summed
$E_{T}$ for the final selection (right)} \label{semifig}
\end{figure*}

A simple and robust event selection was developed,
requesting exactly one tightly isolated muon with $p_{T} > 30$ GeV
and at least four jets with $E_{T} > 65$ GeV for the leading jet and
$E_{T} > 40$ GeV for the other jets. For 10 $pb^{-1}$  of integrated
luminosity, 128 signal events are expected, corresponding to a
selection efficiency of 10.3$\%$, together with 25 other $t\bar{t}$
final state events, as well as 45 (7) W+jets (Z+jets) events. The
number of QCD background events is determined as 11 with a large
uncertainty. Figure~\ref{semifig} shows the jet multiplicity
distribution for events passing the final selection except the
requirement of $N_{jets} \geq{4}$ and the invariant mass of the
three jets with the highest vectorially summed $E_{T}$ for the final
selection.

\subsection{Expectations for observation of top quark pair production in the dilepton final state with the first 10 $pb^{-1}$ of CMS data and measurements of the $t\bar{t}$ cross section with the fist 100 $pb^{-1}$} \label{dilsec}
\begin{figure*}[t] \centering
\includegraphics[width=45mm]{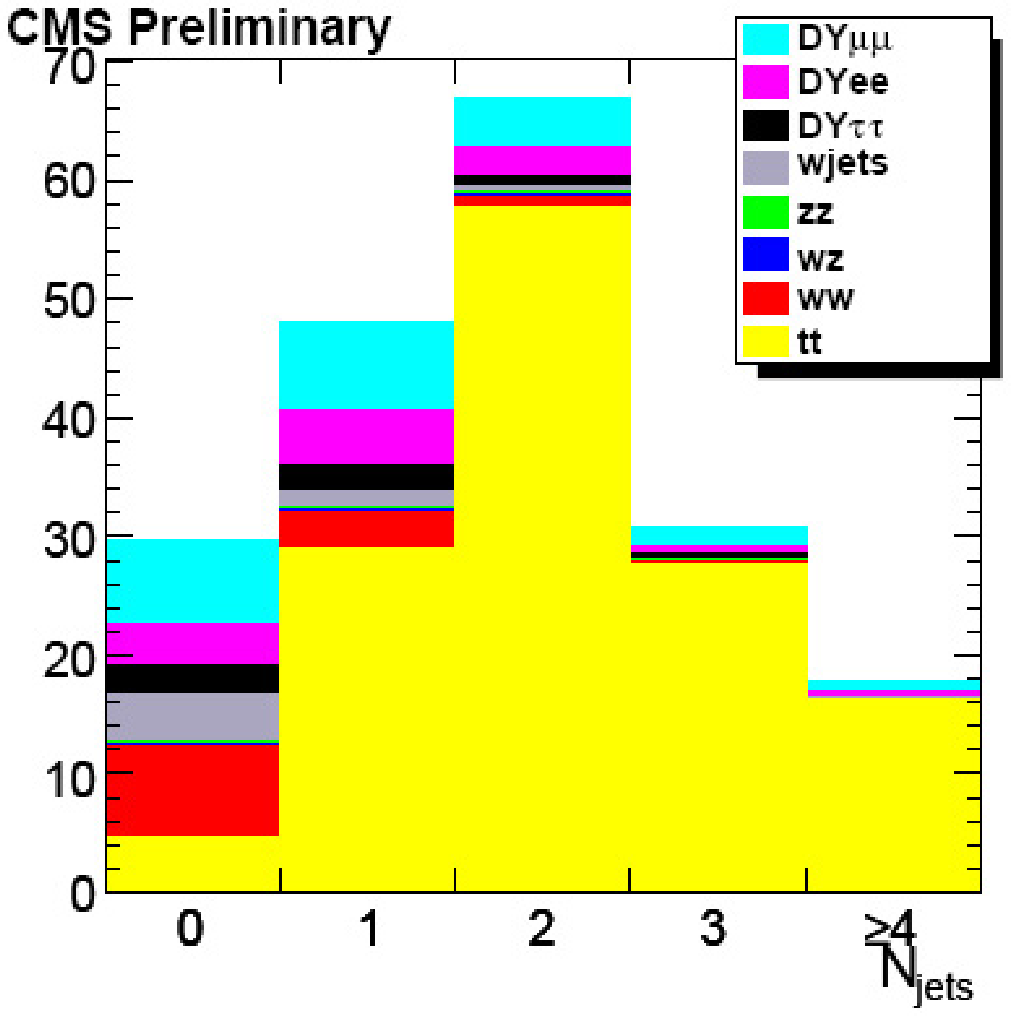}
\includegraphics[width=70mm]{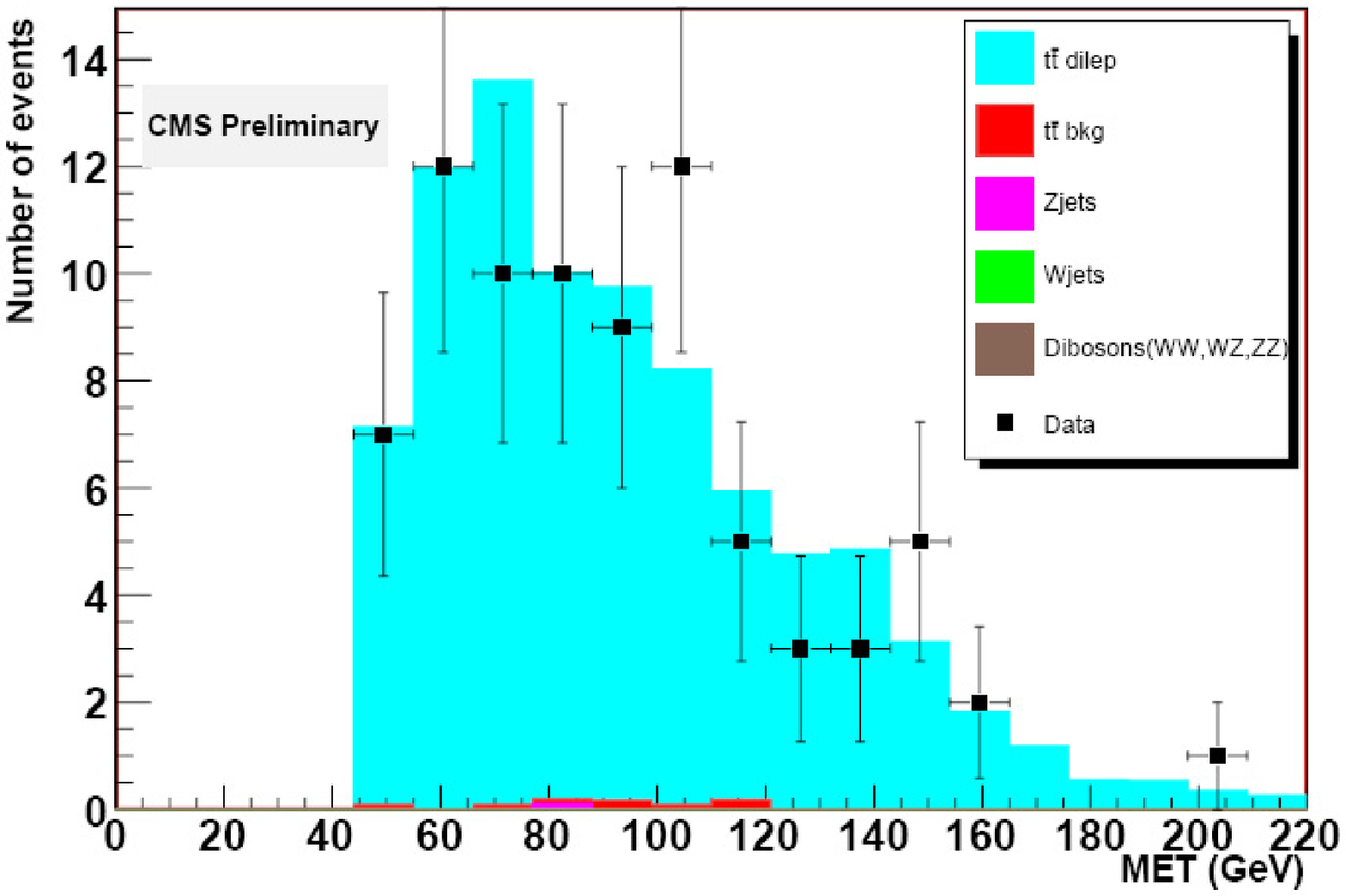}
\caption{The expected number of dilepton events in the first 10
$pb^{-1}$ passing the final selection except the requirement of
$N_{jets} \geq{2}$ (left); The missing transverse energy for the
final selected events in $e\mu$ channel for a total integrated
luminosity of 100 $pb^{-1}$ (right)} \label{dilfig}
\end{figure*}

A clear signal stands out in the first 10 $pb^{-1}$ of CMS data in
events selected with two high momentum leptons, missing transverse
energy, and at least two jets with expected signal-to-noise ratio of
7 to 1 in all events combined and 25 to 1 in electron-muon final
state alone. The most important background comes from Z+jets events.
For an integrated luminosity of 100 $pb^{-1}$ the detector is
assumed to be calibrated and aligned with an amount of data
corresponding to 10 $pb^{-1}$. An additional cut on a robust b-jet
identification discriminator based on a track counting algorithm
leads to an almost background free event selection where about 160
signal events are expected. For this scenario the cross-section can
be extracted by means of a robust event counting method and the
associated statistical uncertainty is found to be $8\%$.
Figure~\ref{dilfig} shows the expected number of dilepton events in
the first 10 $pb^{-1}$ passing the final selection except the
requirement of $N_{jets} \geq{2}$ and the missing transverse energy
for the final selected events in $e\mu$ channel for a total
integrated luminosity of 100 $pb^{-1}$.

\subsection{Toward the measurement of the cross section in the etau and mutau dilepton channel} \label{tausec}
 For this study taus are defined as tracks
approximately collinear with a jet $(\Delta R(track, jet-axis) <
0.1)$ and isolated of other tracks in the cone $(\Delta R < 0.45)$.
Using an event selection strategy similar to the ones described in
2.2 or 2.3, approximately 86 events are found in a simulated event
sample corresponding to the first 100 $pb^{-1}$ of integrated
luminosity. A signal-to-background ratio $S/B \approx 0.4$ is found
for 1-prong tau decays, where the most important background comes
from the W+jets events.

\section{TOP QUARK DECAYS AS CALIBRATION TOOLS}

\subsection{Measurement of jet energy scale corrections using top quark events}
A least-square kinematic fit using Lagrange multipliers can be
applied to enforce the mass constraint in the reconstructed $pp
\rightarrow t\bar{t} \rightarrow q\bar{q}b\mu\nu_{\mu}\bar{b}$
events. Both the W boson mass and the top quark mass in the hadronic
top quark decay, $t \rightarrow Wbt \rightarrow q\bar{q}b$, are
constrained to agree with their measured values. Residual
corrections are estimated on the energy scale of the jets arising
from both the light quarks in the W boson decay and the heavier
bottom quark in the top quark decay. Utilizing the first integrated
luminosity of 100 $pb^{-1}$ of CMS data, an uncertainty smaller than
1$\%$ can be obtained on the jet energy scale for both light and
heavy jets. More details can be found in~\cite{refjet}

\section{CONCLUSIONS}
Observability of top quark signal for different final states of
$t\bar{t}$ production has been presented. It has been shown that the
top quark signature can be established with the first 10 $pb^{-1}$
in both the semileptonic muon channel, where about 128 signal events
are expected with a signal-to-noise ratio of about 1.5, and in the
dilepton channel, where about 80 signal events are expected in the 3
possible final states with a combined signal-to-noise ratio of 7 to
1. In addition the semileptonic muon channel final state can be used
to obtain the jet energy scale corrections for both light and heavy
jets with an uncertainty smaller than 1$\%$ for an integrated
luminosity of 100 $pb^{-1}$.

\end{document}